\documentclass[conference]{IEEEtran}
\IEEEoverridecommandlockouts
\usepackage{cite}
\usepackage{amsmath,amssymb,amsfonts}
\usepackage{algorithmic}
\usepackage{graphicx}
\usepackage{textcomp}
\usepackage{xcolor}
\usepackage{multirow}
\usepackage{booktabs}

\allowdisplaybreaks[4]
\newcounter{num}
\newcommand{\Rnum}[1]{\setcounter{num}{#1} \Roman{num}}

\def\BibTeX{{\rm B\kern-.05em{\sc i\kern-.025em b}\kern-.08em
    T\kern-.1667em\lower.7ex\hbox{E}\kern-.125emX}}
\begin{document}

\title{Inter-subject Variance Transfer Learning for EMG Pattern Classification Based on Bayesian Inference}
\newcommand{\argmax}{\mathop{\rm arg\ max}\limits}

\author{\IEEEauthorblockN{Seitaro Yoneda and Akira Furui}
\IEEEauthorblockA{
Graduate School of Advanced Science and Engineering, Hiroshima University, Higashi-hiroshima, Japan.\\
Email: \{seitaroyoneda, akirafurui\}@hiroshima-u.ac.jp
}
}

\maketitle

\begin{abstract}
In electromyogram (EMG)-based motion recognition, a subject-specific classifier is typically trained with sufficient labeled data.
However, this process demands extensive data collection over extended periods, burdening the subject.
To address this, utilizing information from pre-training on multiple subjects for the training of the target subject could be beneficial.
This paper proposes an inter-subject variance transfer learning method based on a Bayesian approach. 
This method is founded on the simple hypothesis that while the means of EMG features vary greatly across subjects, their variances may exhibit similar patterns.
Our approach transfers variance information, acquired through pre-training on multiple source subjects, to a target subject within a Bayesian updating framework, thereby allowing accurate classification using limited target calibration data. 
A coefficient was also introduced to adjust the amount of information transferred for efficient transfer learning.
Experimental evaluations using two EMG datasets demonstrated the effectiveness of our variance transfer strategy and its superiority compared to existing methods.
\end{abstract}


\section{Introduction}

Electromyogram (EMG) signals are electrical activities that can be measured from the skin surface, and their characteristics change depending on the motion performed.
Machine learning leverages EMG signals in various fields, such as myoelectric prostheses, by estimating motion intentions~\cite{Tsoli2011-jv,Rezazadeh2012-ol}.

In machine learning-based EMG pattern recognition, unique classifiers are commonly trained using data collected individually from each subject~\cite{Zhang2022-ld,Matsubara2011-pc}.
This is due to substantial individual differences exist in EMG signals, which can be attributed to variations in subcutaneous fat distribution, muscle fiber diameter, and the manner in which force is exerted~\cite{Matsubara2011-pc,Wu2023-bg}.
However, this approach requires the collection of extensive training data over multiple trials to ensure adequate recognition performance, which poses a heavy burden on subjects and presents a major challenge for practical applications.

Inter-subject transfer learning has been gaining attention as a method for achieving high recognition performance while reducing data collection costs for a new subject. 
This approach utilizes the knowledge from pre-training with multiple subjects (\textit{source} subjects) to train a classifier for a new subject (\textit{target} subject), anticipating that in such a scenario, only minimal labeled calibration data will be collected from the target subject. 
Recently, transfer learning based on deep neural networks has been widely used~\cite{Zou2021-zy,Kim2020-ae}; however, issues related to computational cost and overfitting have been identified~\cite{Bao2021-zq}.

For applications on edge devices with limited computational resources, simpler, non-deep classification models are preferable for inter-subject transfer learning. 
Traditionally, a framework of transfer learning is introduced for non-deep classification models~\cite{Vidovic2016-bm,Kanoga2021-ve,Vidovic2014-jq}. 
For instance, in adaptive linear discriminant analysis (LDA)~\cite{Vidovic2016-bm}, model parameters (mean and covariance matrix) from both the target and source are combined.
The weight of the combination is then empirically adjusted to determine the amount of information transferred from source to target, to achieve effective recognition.

As a non-deep strategy, the authors have explored transfer learning using simple Bayesian models in EMG classification~\cite{Yoneda2023-lj}.
In Bayesian transfer learning, the transfer of knowledge is interpreted as a process, whereby the prior distribution is transformed into the posterior distribution via the likelihood.
The amount of transfer from source to target is determined based on the uncertainty of priors. 
We demonstrated the viability of this approach for intra-subject, inter-trial transfer learning~\cite{Yoneda2023-lj}.
However, its applicability and success in inter-subject contexts requires further investigation.

\begin{figure}[t]
  \centering
    \includegraphics[width=\hsize]{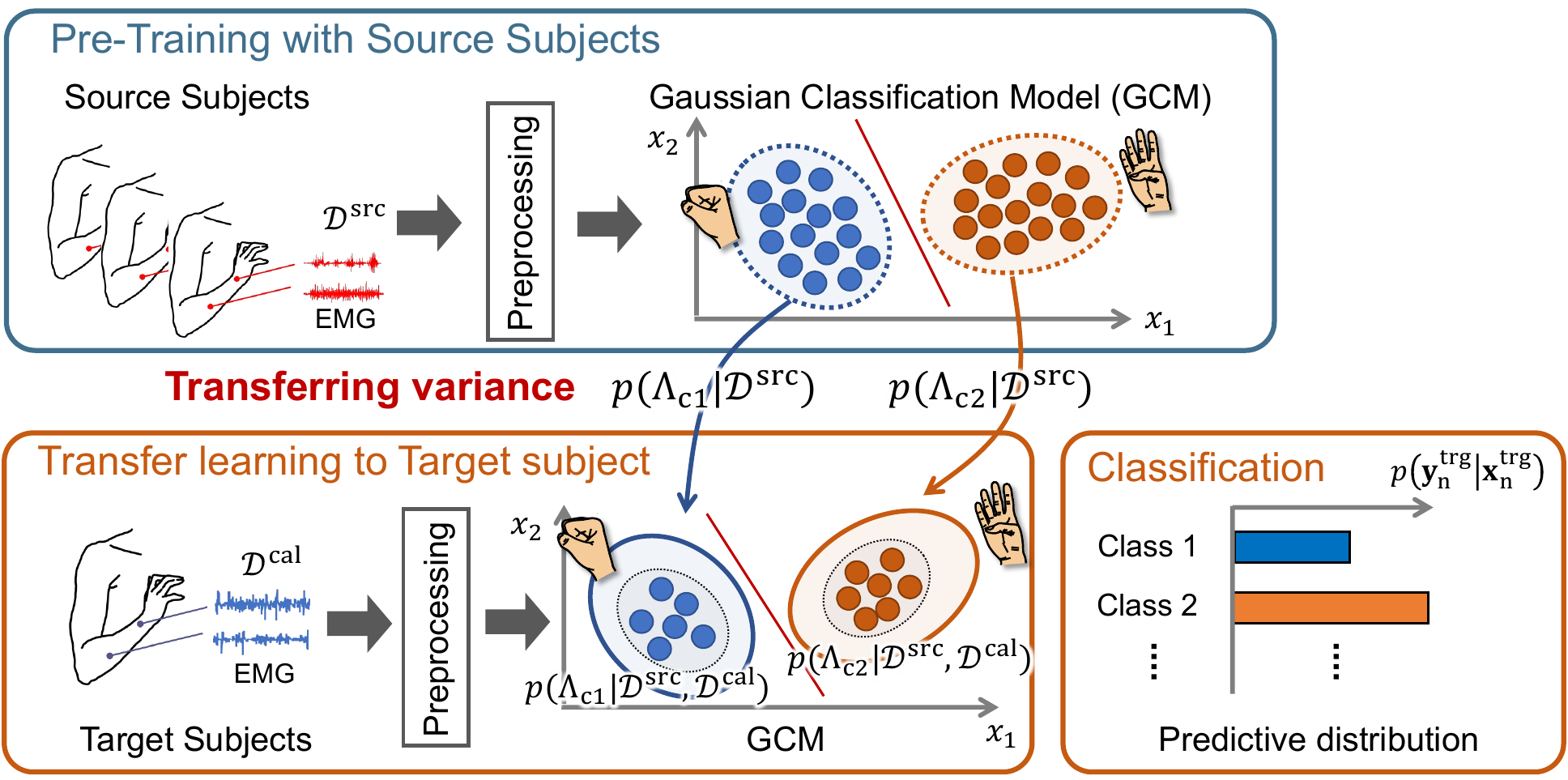}
  \caption{Overview of the proposed method}
   \label{fig :overview}
\end{figure}

This paper presents a Bayesian approach to inter-subject transfer learning in EMG classification, employing a simple Gaussian classification model (GCM).
Fig.~\ref{fig :overview} presents an overview of the proposed method. 
The method initially computes the posterior distributions of GCM parameters using EMG signals from multiple source subjects.
These distributions are then updated using calibration data from a target subject. 
The proposed method shares the precision matrix among subjects, thereby transferring only variance information from source to target subjects.
The amount of transferred information is determined based on the uncertainty of the pre-trained posterior distribution on the source subjects.

\section{Proposed method}

\subsection{Problem Setting}
Consider the $D$-dimensional EMG feature vector at data point $n$, denoted as $\mathbf{x}_{n}\in \mathbb{R}^{D}$, and its corresponding motion class label $\mathbf{y}_n\in\{0,1\}^{C}$, where $C$ is the number of classes. 
In inter-subject transfer learning for EMG signals, the information learned from the data of multiple source subjects is utilized during the training of a classification model for a target subject, to enhance classification performance for the target. 
Let us consider the pre-training of a model with a labeled training dataset obtained from a source subject $s$ as $\mathcal{D}^{s} = \{(\mathbf{x}^{s}_{n},\mathbf{y}^{s}_{n})\}_{n=1}^{N^{s}}$, where $N^{s}$ is the number of data points in the source subject. 
The complete dataset across all source subjects is $\mathcal{D}^{\mathrm{src}}=\{\mathcal{D}^{s}\}_{s=1}^{S}$, with a total sample size of $N^{\mathrm{src}}=\sum_{s}N^{s}$, which is assumed to be sufficiently large.
The target subject provides a small set of labeled data (referred to as calibration data), denoted as $\mathcal{D}^{\mathrm{cal}}=\{(\mathbf{x}^{\mathrm{cal}}_{n},\mathbf{y}^{\mathrm{cal}}_{n})\}_{n=1}^{N^{\mathrm{cal}}}$. 
Through transfer learning, the ultimate goal is to maximize the classification performance on the test data $\mathbf{x}^{\mathrm{trg}}_{n}$ for the target subject.

\subsection{Gaussian Classification Model (GCM)}
A GCM represents the relationship between EMG signals and the corresponding class labels.
The observed model of the EMG signal $\mathbf{x}_{n}$ for class $c \in \{1,\dots C\}$ is expressed via the following Gaussian distribution:
\begin{align}
    p(\mathbf{x}_n|\boldsymbol{\mu}_{c},\mathbf{\Lambda}_{c})=\mathcal{N}(\mathbf{x}_n|\boldsymbol{\mu}_{c},\mathbf{\Lambda}_{c}^{-1}), \label{Gaussian}
\end{align}
where $\boldsymbol{\mu}_{c}\in\mathbb{R}^{D}$ and $\boldsymbol{\Lambda}_{c}\in\mathbb{R}^{D\times D}$ represent the mean vector and precision matrix (the inverse of the covariance matrix) of each class, respectively.
We use one-hot representation for the motion class label $\mathbf{y}_{n}=\{y_{n,c}\}$, where $y_{n,c}=1$ for a specific $c$ indicates the selection of the $c$-th class.
Hence, the distribution of $\mathbf{y}_n$ can be represented using the following categorical distribution:
\begin{align}
    p(\mathbf{y}_{n}|\boldsymbol{\pi})=\text{Cat}(\mathbf{y}_{n}|\boldsymbol{\pi}) 
    =\prod_{c=1}^{C}{\pi}_{c}^{y_{n,c}}, \label{categorical}
\end{align}
where $\boldsymbol{\pi}=\{\pi_c\}$ denotes the mixing coefficients ($\pi_{c}\in[0,1]$ and $\sum_{c=1}^{C}\pi_{c}=1$).

A prior distribution is introduced for each model parameter for Bayesian treatment.
For computational convenience, we set the Gaussian-Wishart and Dirichlet distributions, which are conjugate priors for $\{\boldsymbol{\mu}_c, \boldsymbol{\Lambda}_c\}$, and $\boldsymbol{\pi}$, as follows:
\begin{align}
    p(\boldsymbol{\mu}_{c},\mathbf{\Lambda}_{c}) &= p(\boldsymbol{\mu}_{c}|,\mathbf{\Lambda}_{c})p(\mathbf{\Lambda}_{c})\\
    &=\mathcal{N}(\boldsymbol{\mu}_{c}|\mathbf{m}_{0},(\beta_{0} \mathbf{\Lambda}_{c})^{-1})\mathcal{W}(\mathbf{\Lambda}_{c}|\mathbf{\nu}_{0},\mathbf{W}_{0}), \label{gauswhisert}\\
    p(\boldsymbol{\pi})&=\mathrm{Dir}(\boldsymbol{\pi}|\boldsymbol{\alpha}_0), \label{dirikure} 
\end{align}
where $\mathbf{m}_0\in\mathbb{R}^{D}$, $\beta_0\in\mathbb{R}^{+}$, $\nu > D - 1$, $\mathbf{W}_0\in\mathbb{R}^{D\times D}$, and $\boldsymbol{\alpha}_0 = \{\alpha_0\}^C$ are the prior hyperparameters.

\begin{figure}[t]
    \centering
      \includegraphics[width=\hsize]{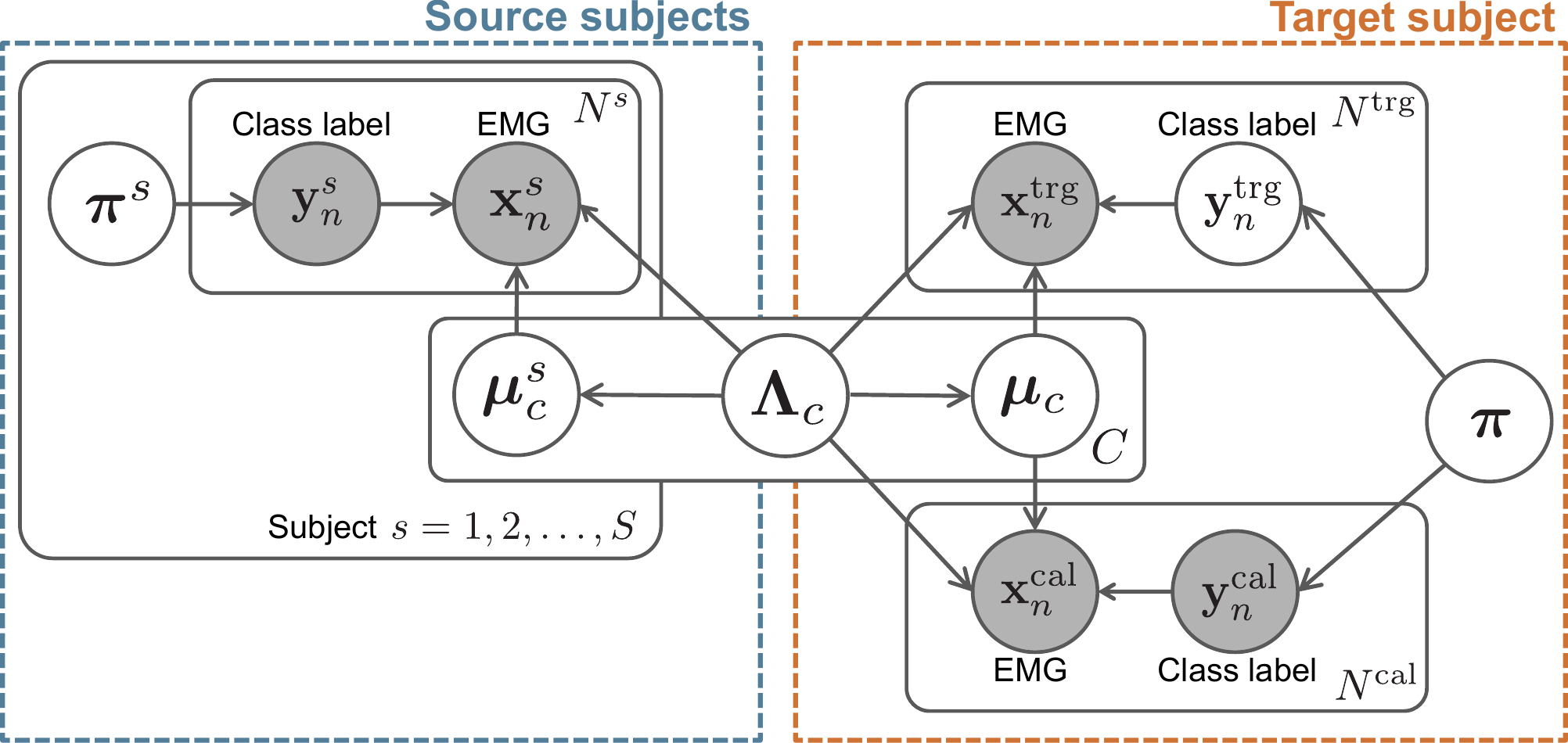}
    \caption{Graphical model of GCM for inter-subject classification. The gray and white nodes represent the observed and unobserved variables, respectively.}
     \label{fig :Graphicalmodel}
\end{figure}

\subsection{Inter-Subject Transfer Learning}

In transfer learning based on structured Bayesian models, the issue reduces to whether to share parameters among subjects or to treat them independently.
EMG feature vectors vary significantly across subjects due to individual biological differences, thereby affecting the mean values of the features.
By contrast, feature variance (the variability over time) may display common trends across subjects.
With these simple assumptions, we set subject-specific means and a shared variance in GCM.
Fig.~\ref{fig :Graphicalmodel} shows the graphical model of the extended GCM.
The training steps of the model are divided into pre-training for source subjects and transfer learning for the target subject.

\begin{table*}[t]
    \caption{Dataset information}
    \label{table: dataset information}
    \centering
    \begin{tabular}{lllllll}
       \toprule
       Dataset & \# Motions ($C$) & \# Electrodes ($D$) & Sampling frequency & \# Subjects & Calibration trials & Test trials \\
       \midrule
       \Rnum{1}  & $6$ &$4$& $2000$ Hz & $6$  & $1$ & $2,3,4,5,6$\\
       \Rnum{2} & $8$ &$8$& $200$ Hz & $25$ & $1$ & $2,3,4$\\
       \bottomrule
    \end{tabular}
\end{table*}

\subsubsection{Pre-training with source subjects}
The posterior distributions of the source subject $s$ for each class can be computed by applying  Bayes' theorem as follows:
\begin{align}
p(\boldsymbol{\mu}_{c}^{s},\mathbf{\Lambda}_{c}|\mathcal{D}^\mathrm{src}) &\propto p(\mathcal{D}^\mathrm{src} | \boldsymbol{\mu}_{c}^{s}, \boldsymbol{\Lambda}_{c} ) p (\boldsymbol{\mu}_{c}^{s} | \boldsymbol{\Lambda}_{c})p (\boldsymbol{\Lambda}_{c}),\\
p(\boldsymbol{\pi}^{s}|\mathcal{D}^{\mathrm{src}})&\propto p(\mathcal{D}^{\mathrm{src}}|\boldsymbol{\pi}^{s})p(\boldsymbol{\pi}^{s}).
\end{align}
From conjugacy, each posterior distribution is attributed to the following Gaussian-Wishart and Dirichlet distributions, which are of the same type as the priors:
\begin{align}
p(\boldsymbol{\mu}_{c}^{s},\mathbf{\Lambda}_{c}|\mathcal{D}^\mathrm{src}) &=\mathcal{N}(\boldsymbol{\mu}_{c}^{s}|\mathbf{m}_{c}^{s},(\beta_{c}^{s} \mathbf{\Lambda}_{c})^{-1})\mathcal{W}(\mathbf{\Lambda}_{c}|\mathbf{\nu}_{c}^\mathrm{src},\mathbf{W}_{c}^\mathrm{src})\label{mu,lamda},\\
\quad p(\boldsymbol{\pi}^{s}|\mathcal{D}^\mathrm{src})&=\mathrm{Dir}(\boldsymbol{\pi}^{s}|\boldsymbol{\alpha}^{s})\label{pi},
\end{align}
where $\beta_{c}^{s},\mathbf{m}_{c}^{s},\nu_{c}^\mathrm{src},\mathbf{W}_{c}^\mathrm{src}$, and $\boldsymbol{\alpha}^{s} = \{{\alpha}_{c}^{s}\}_{c=1}^C$ are the posterior hyperparameters for the source subjects.
Note that the hyperparameters for the subject-specific mean carry a subject index $s$, whereas for the shared variance (precision) across subjects, the superscript ``$\mathrm{src}$'' denotes its commonality.

In Bayesian transfer learning, the amount of transferred information is determined by the uncertainty of the distribution, which is inversely related to data quantity. 
With more source subjects or larger samples, there is the risk of transferring excessive information to the target. 
Accordingly, we introduce a weight coefficient $w^s \in \mathbb{R}^+$ to control the distribution uncertainty and adjust the impact of source data. 
Using this coefficient, the modified posterior hyperparameters for the source subjects can be calculated as follows:
\begin{align}
\beta_{c}^{s}&=\sum_{n=1}^{N^{s}}w^sy^{s}_{n,c}+\beta_{0},\\ \mathbf{m}_{c}^{s}&=\frac{\sum_{n=1}^{N^{s}}w^sy^{s}_{n,c}\mathbf{x}^{s}_{n}+\beta_{0}\mathbf{m}_{0}}{{\beta}_{c}^{s}},\\
\nu_{c}^\mathrm{src}&=\frac{1}{S}\sum_{s=1}^{S}\sum_{n=1}^{N^{s}}w^sy^{s}_{n,c}+\nu_{0}\label{fl-nu},  \\
{(\mathbf{W}_{c}^\mathrm{src})}^{-1}&=\frac{1}{S}\sum_{s=1}^{S}\left( \sum_{n=1}^{N^{s}}w^sy^{s}_{n,c}\mathbf{x}^{s}_{n}{\mathbf{x}^{s}_{n}}^{\top}+\beta_{0}\mathbf{m}_{0}\mathbf{m}_{0}^{\top}\notag \right.\\
& \left. \quad \quad \quad \quad \quad -\beta_{c}^{s}\mathbf{m}_{c}^{s}{\mathbf{m}_{c}^{s}}^{\top}\right)+\mathbf{W}_{0}^{-1},\label{fl-w}  \\
\alpha_{c}^{s}&=\sum_{n=1}^{N^{s}}y^{s}_{n,c}+\alpha_{0}.
\end{align}
Note that a larger $w^{s}$ reduces distribution uncertainty, increasing the amount of information transferred from source to target.

\subsubsection{Transfer learning to target subject}
The model for the target subject is trained with transferred information from pre-training on source subjects.
The posterior distributions for the target subject are computed based on both $\mathcal{D}^{\mathrm{src}}$ and $\mathcal{D}^{\mathrm{cal}}$, as follows:
\begin{align}
p(\boldsymbol{\mu}_c,\mathbf{\Lambda}_ {c}|\mathcal{D}^{\text{cal}},\mathcal{D}^{\mathrm{src}})&\propto p(\mathcal{D}^{\mathrm{cal}}|\boldsymbol{\mu}_{c},\mathbf{\Lambda}_{c})p(\boldsymbol{\mu}_{c}|\mathbf{\Lambda}_{c})p(\mathbf{\Lambda}_c|\mathcal{D}^{\mathrm{src}}),\\
p(\boldsymbol{\pi}|\mathcal{D}^\mathrm{cal})&\propto p(\mathcal{D}^{\mathrm{cal}}|\boldsymbol{\pi})p(\boldsymbol{\pi}).
\end{align}
Note that only the precision matrix $\mathbf{\Lambda}_ {c}$ is shared between source and target, conditioned on $\mathcal{D}^\mathrm{src}$.
Similar to (\ref{mu,lamda}) and (\ref{pi}), each posterior distribution can be calculated using conjugation as follows:
\begin{align}    
p(\boldsymbol{\mu}_{c},\mathbf{\Lambda}_{c}|&\mathcal{D}^\mathrm{cal},\mathcal{D}^\mathrm{src}) \notag \\
&=\mathcal{N}(\boldsymbol{\mu}_{c}|\mathbf{m}_{c},(\beta_{c}\mathbf{\Lambda}_{c})^{-1})  \mathcal{W}(\mathbf{\Lambda}_{c}|\mathbf{\nu}_{c},\mathbf{W}_{c}),\\
p(\boldsymbol{\pi}|\mathcal{D}^{\text{cal}})&=\mathrm{Dir}(\boldsymbol{\pi}|\boldsymbol{\alpha}),\quad\quad\quad \quad\quad\quad\quad \quad
\end{align}
where $\beta_{c},\mathbf{m}_{c},\nu_{c},\mathbf{W}_{c},\boldsymbol{\alpha} = \{\alpha_{c}\}$ are the posterior hyperparameters for the target subject and defined as follows:
\begin{align}
\beta_{c}&=\sum_{n=1}^{N^{\mathrm{cal}}}y^{\mathrm{cal}}_{n,c}+\beta_{0},\\ \mathbf{m}_{c}&=\frac{\sum_{n=1}^{N^{\mathrm{cal}}}y^{\mathrm{cal}}_{n,c}\mathbf{x}^{\mathrm{cal}}_{n}+\beta_{0}\mathbf{m}_{0}}{\beta_{c}},\\
\nu_{c}&=\sum_{n=1}^{N^{\mathrm{cal}}}y^{\mathrm{cal}}_{n,c}+\nu_{c}^{\mathrm{src}},\label{target-nu} \\    
\mathbf{W}_{c}^{-1}&=\sum_{n=1}^{N^{\mathrm{cal}}}y^{\mathrm{cal}}_{n,c}\mathbf{x}^{\mathrm{cal}}_{n}{\mathbf{x}^{\mathrm{cal}}_{n}}^{\top}+\beta_{0}\mathbf{m}_{0}\mathbf{m}_{0}^{\top}\notag\\
& \quad \quad \quad \quad \quad -\beta_{c}\mathbf{m}_{c}\mathbf{m}_{c}^{\top}+{(\mathbf{W}_{c}^\mathrm{src})}^{-1},\label{target-w}\\
\alpha_{c}&=\sum_{n=1}^{N^{\mathrm{cal}}}y^{\mathrm{cal}}_{n,c}+\alpha_{0}.
\end{align}
From (\ref{target-nu}) and (\ref{target-w}), the calculation of the posterior hyperparameters for $\mathbf{\Lambda}_{c}$ utilizes hyperparameters determined by the source subjects. 
This allows transfer of variance information from source to target.

\subsection{Motion class prediction}
To perform motion recognition on new target subject test data $\mathbf{x}_{n}^{\mathrm{trg}}$, we compute the following predictive distribution:
\begin{align}
  p(\mathbf{y}_{n}^{\mathrm{trg}} | &\mathbf{x}_{n}^{\mathrm{trg}}, \mathcal{D}_\mathrm{cal}, \mathcal{D}_\mathrm{src})\notag \\ 
  &= \int p(\mathbf{y}_{n}^{\mathrm{trg}} | \mathbf{x}_{n}^{\mathrm{trg}}, \boldsymbol{\theta}_{c}) p(\boldsymbol{\theta}_{c} | \mathcal{D}_{\mathrm{cal}}, \mathcal{D}_{\mathrm{src}}) \mathrm{d} \boldsymbol{\theta}_{c}, \label{kankeishiki}
\end{align}
where $\boldsymbol{\theta}_{c}=\{\boldsymbol{\mu}_{c}, \boldsymbol{\Lambda}_{c},\boldsymbol{\pi}\}$ represents the set of model parameters.
The class prediction probabilities defined by the predicted distribution are then calculated, and the class with the highest probability is determined to be the predicted motion class:
\begin{align}
  \hat{c}_n=\argmax_{c}p(\mathbf{y}_{n,c}^{\mathrm{trg}}=1 | &\mathbf{x}_{n}^{\mathrm{trg}}, \mathcal{D}_\mathrm{cal}, \mathcal{D}_\mathrm{src}).
\end{align}

\section{Experiments}
\subsection{Methods}
To evaluate the effectiveness of the proposed inter-subject transfer learning, EMG pattern classification experiments were conducted using two EMG datasets. 
Table~\ref{table: dataset information} summarizes the information for each dataset, which containes multi-channel EMG signals from several subjects over multiple trials. 
Datasets I and II are obtained from~\cite{Yoneda2023-lj} and \cite{Kanoga2021-ve}, respectively. 
For feature extraction, each dataset was subjected to full-wave rectification processing and smoothing using a second-order Butterworth low-pass filter with a cutoff frequency of $1$ Hz.
One subject was designated as the target subject, and the rest as source subjects, with validation performed on all possible combinations.
All trials from source subjects were used as the pre-training set $\mathcal{D}^\mathrm{src}$. 
For the target subject, only the first trial served as the calibration set $\mathcal{D}^\mathrm{cal}$, and the remaining trials served as the test set for accuracy evaluation.

The prior hyperparameters for GCM were set as $\mathbf{m}_{0}=\mathbf{0}$, $\beta_0=1$, $\nu_0 = D + 1$, $\boldsymbol{\alpha}_0 = 1$, and $\mathbf{W}_0=\mathbf{I}$ ($\mathbf{I}$ as the identity matrix). 
The weight coefficient $w^s$ introduced during pre-training, is an important parameter for adjusting the amount of information transferred from source to target.
In our experiments, we tentatively set $w^s$ as follows: 
\begin{align}
  w^s=\frac{N^{\mathrm{cal}}}{N^s}
\end{align}
so that the ratio of the amount of transferred data is equal to that of the calibration data.
This setting assumes that if the transfer amount is too high, the information from source subjects may dominate, hindering the appropriate incorporation of the target subject information into the learning process.

To investigate the impact of varying transferred information, we performed an ablation study for the proposed method.
We also compared the performance of the proposed method with the following two existing methods:
\begin{itemize}
    \item \textbf{Adaptive LDA}~\cite{Vidovic2016-bm} is a classification method that introduces transfer learning into LDA. 
    In this method, the model parameters, ${\boldsymbol{\mu}}_c$ and ${\mathbf{\Sigma}}$, are determined as follows:
    \begin{align}
    {\boldsymbol{\mu}}_c &= \tau \boldsymbol{\mu}_c^\mathrm{cal} + (1-\tau) \boldsymbol{\mu}_c^\mathrm{src},\\
    {\mathbf{\Sigma}} &= \lambda \mathbf{\Sigma}^\mathrm{cal} + (1-\lambda) \mathbf{\Sigma}^\mathrm{src},
    \label{eq:alda}
    \end{align}
    where $\boldsymbol{\mu}_c^\mathrm{src}$ and $\mathbf{\Sigma}^\mathrm{src}$ are the class-specific mean and shared covariance matrix calculated from the source dataset, respectively, whereas $\boldsymbol{\mu}_c^\mathrm{cal}$ and $\mathbf{\Sigma}^\mathrm{cal}$ are those from the target subject's calibration dataset.
    The hyperparameters $\{\tau, \lambda\}$ control the amount of transferred information and were tuned using $11\times 11$ grid search (ranging from 0 to 1) through two-fold cross-validation on the calibration dataset.

    \item \textbf{Adaptive quadratic discriminant analysis (Adaptive QDA)}~\cite{Vidovic2014-jq} introduces transfer learning into QDA, which is an extension of LDA, where each class is given a class-specific covariance matrix $\mathbf{\Sigma}_c$.
    Unlike LDA, QDA can learn quadratic decision boundaries. 
    The model parameters, ${\boldsymbol{\mu}}_c$ and ${\mathbf{\Sigma}}_c$, can be determined by replacing $\boldsymbol{\Sigma}^\mathrm{cal}$ and $\boldsymbol{\Sigma}^\mathrm{src}$ in (\ref{eq:alda}) with $\boldsymbol{\Sigma}_c^\mathrm{cal}$ and $\boldsymbol{\Sigma}_c^\mathrm{src}$, respectively.
    Tuning of the hyperparameters was performed in the same way as for the adaptive LDA.
\end{itemize}
In both the ablation study and the comparison with existing methods, we evaluated the effect of calibration data size on accuracy using the entire set and only the first 25\%.
Apart from the above evaluations, the sensitivity of the proposed method to the coefficient $w^s$ was also explored.

\subsection{Results}
In the ablation study, we calculated the classification accuracy for different transfer scenarios: transferring only variance information (ours), only mean information, both mean and variance information, and not transferring any information. The results are presented in 
Table \ref{table:baselines}.
Regardless of the dataset and the percentage of calibration data used, the proposed method of transferring only variance information demonstrated the best performance.
Table~\ref{table:existing} shows the results of the comparison with existing methods. 
The proposed method outperformed the existing methods in all cases.

\begin{figure}[t]
  \begin{minipage}[b]{\linewidth}
    \centering
    \includegraphics[width=\hsize]{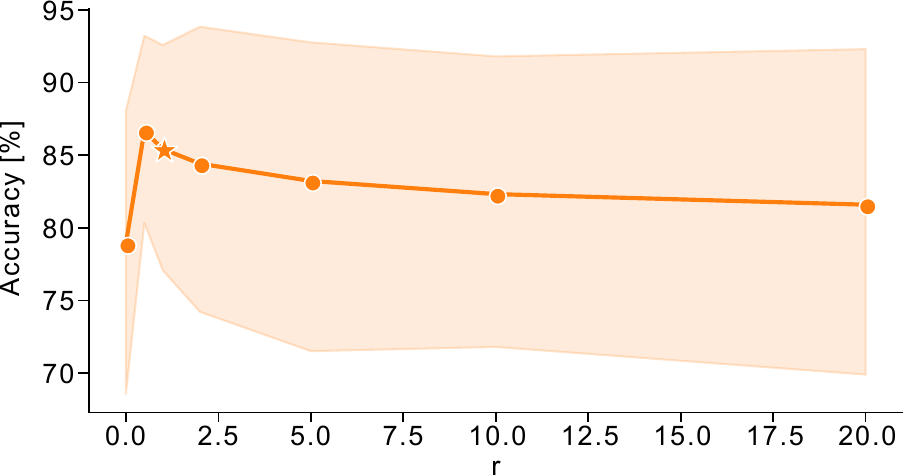}
    \centering{\footnotesize{(a) Dataset\Rnum{1}}}
  \end{minipage}\\
  \begin{minipage}[b]{\linewidth}
    \centering
    \includegraphics[width=\hsize]{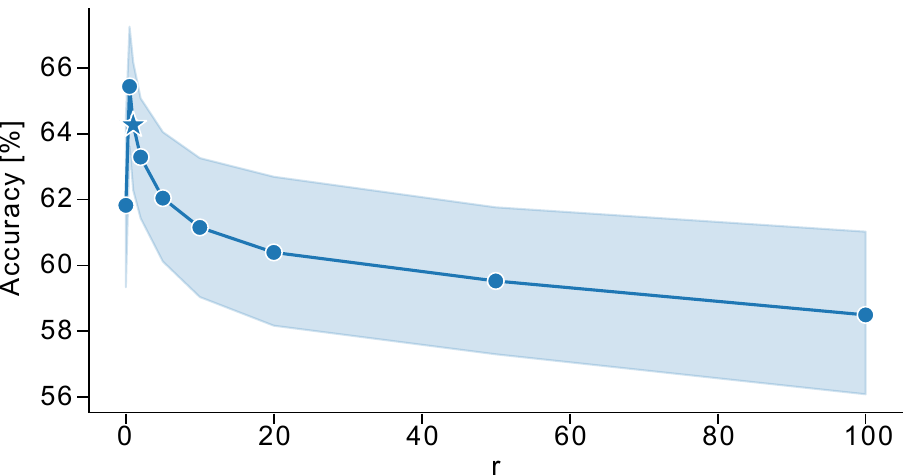}
    \centering{\footnotesize{(b) Dataset\Rnum{2}}}
  \end{minipage}
  \caption{Average classification accuracy with changing $r$ for each dataset. Note that the maximum value of $r$ varies across datasets due to differences in the number of source subjects and data length of them. Each point and the shaded areas represent average accuracy and $95\%$ confidence interval respectively, for all participants. Stars indicate an accuracy corresponding to $r=1$.}
  \label{fig:weight}
\end{figure}

The effect of modulating the transfer amount by varying the coefficient $w^s$ on performance was also explored.
As the number of source subjects, $S$, varied across datasets, we expressed $w^s$ as a ratio $r$ of the volumes of source to calibration data, as follows: 
\begin{align}
  r=\frac{\sum_{s=1}^{S}w^sN^s}{S N^{\mathrm{cal}}}.
\end{align}
Performance was assessed at different $r$ values ($r = 0$, $0.5$, $1$, $2$, $5$, $10$, $20$, $50$, $100$), and the results are shown in Fig.~\ref{fig:weight}.
Note that in previous ablation and comparative studies, the default $w^s$ corresponded to $r = 1.0$.
In both datasets, average accuracy peaked around $0.5$--$1.0$ and then gradually decreased as $r$ increased.

\begin{table}[t]                     
    \caption{Average classification accuracy (\%) in ablation study}
    \label{table:baselines}
    \centering
     \begin{tabular}{@{}clcc@{}}
      \toprule
      Dataset & Type of transfer & \multicolumn{2}{c}{\% of calibration data used}\\
      \cmidrule{3-4}  
          & & 25\% & 100\%\\
      \midrule          
      \multirow{4}{*}{\Rnum{1}} 
       & No transfer &$73.27\pm18.42$& $78.89\pm12.46$\\
       & Mean \& variance transfer &$61.10\pm22.99$& $69.64\pm20.05$\\
       & Mean transfer &$62.41\pm16.86$& $75.48\pm20.10$\\
       & Variance transfer (\textbf{ours}) & \scalebox{0.94}[1.0]{$\mathbf{75.99}\pm\mathbf{10.69}$}& \scalebox{0.94}[1.0]{$\mathbf{85.51}\pm\mathbf{10.26}$}\\
     \midrule
     \multirow{4}{*}{\Rnum{2}}
       & No transfer &$20.87\pm7.25$& $61.82\pm6.75$\\
       & Mean \& variance transfer &$47.47\pm8.47$& $60.84\pm5.51$\\
       & Mean transfer &$21.27\pm7.81$ & $61.03\pm6.41$\\
       & Variance transfer (\textbf{ours}) & \scalebox{0.94}[1.0]{$\mathbf{48.49}\pm\mathbf{10.04}$}& \scalebox{0.94}[1.0]{$\mathbf{64.27}\pm\mathbf{4.70}$}\\
     \bottomrule
     \end{tabular}
\end{table}

\begin{table}[t]                     
    \caption{Average classification accuracy (\%) for each method}
    \label{table:existing}
    \centering
     \begin{tabular}{@{}clcc@{}}
      \toprule
      Dataset & Method & \multicolumn{2}{c}{\% of calibration data used}\\
      \cmidrule{3-4}  
          & &25\%&100\%\\
      \midrule          
      \multirow{3}{*}{\Rnum{1}} 
       & Adaptive LDA &$73.72\pm6.43$& $84.49\pm4.92$\\
       & Adaptive QDA &$52.62\pm13.29$& $63.62\pm19.66$\\
       & \textbf{Ours} & \scalebox{0.94}[1.0]{$\mathbf{75.99}\pm\mathbf{10.69}$}& \scalebox{0.94}[1.0]{$\mathbf{85.51}\pm\mathbf{10.26}$}\\
     \midrule
     \multirow{3}{*}{\Rnum{2}} 
       & Adaptive LDA &$31.53\pm13.88$& $60.50\pm3.88$\\
       & Adaptive QDA &$22.98\pm9.16$& $57.86\pm6.87$\\
       & \textbf{Ours} & \scalebox{0.94}[1.0]{$\mathbf{48.49}\pm\mathbf{10.04}$}& \scalebox{0.94}[1.0]{$\mathbf{64.27}\pm\mathbf{4.70}$}\\
     \bottomrule
     \end{tabular}
\end{table}

\subsection{Discussion}
The ablation study showed that the proposed method (variance transfer) outperforms other scenarios (Table~\ref{table:baselines}).
With limited calibration data, the computed variance tends to be underestimated compared to the actual variability during prediction.
The proposed method can better represent the spread of the target data distribution by transferring the variance from source subjects containing sufficient variability, to improve classification accuracy.
In contrast, transferring the mean leads to lower accuracy than no-transfer scenario.
This could be because EMG feature means vary largely among individuals due to biomechanical and physiological factors, and thus sharing mean information may negatively impact model training.

In both datasets, the existing methods, adaptive LDA and adaptive QDA, exhibited lower accuracy compared to the proposed method (Table~\ref{table:existing}). 
The discrepancy in accuracy was especially pronounced when using 25\% of the calibration data, suggesting that the existing methods might overfit the calibration data. 
In particular, QDA, being a nonlinear classifier, is more prone to overfitting than LDA~\cite{Krasoulis2017-dr}. 
By contrast, the proposed Bayesian method inherently includes a form of regularization through the prior distribution.
Additionally, the introduction of weight coefficient $w^s$ allows for adjusting the uncertainty of the distributions (and thus the transferred information) based on the amount of calibration data, leading to higher classification accuracy.

We found that there is an optimal value for the amount of transferred information controlled by the coefficient $w^s$ (Fig.~\ref{fig:weight}); setting it too high leads to excessive transfer of source subject information, resulting in reduced accuracy.
In the performance comparison experiments, we tentatively set $w^s$ such that $r = 1$, which, coincidentally, was close to the average optimal value.
However, as indicated by the error range in the figure, the accuracy variation trends with $r$ significantly differ among individuals.
Future improvements in accuracy can be achieved by introducing theoretical or computational methods for an efficient determination of the optimal $r$.

\section{Conclusion}
In this paper, we proposed an inter-subject GCM-based transfer learning method.
The proposed method can achieve high accuracy with limited labeled training or calibration data by transferring the variance information acquired from the pre-training of source subjects to the target subject.
In addition, by introducing a coefficient, the uncertainty of the pre-trained posterior distributions was adjusted to control the amount of transferred information, thereby preventing accuracy decline from excessive transfer. 

The efficacy of the proposed method was evaluated using two EMG datasets.
The results indicate that the proposed method outperforms other methods in both datasets, demonstrating the effectiveness of transferring variance information in inter-subject transfer learning.
In the future, we will introduce a framework to adjust the optimal amount of transferred information for each subject and improve accuracy through more complex model structures.

\section*{Acknowledgment}
This work was partially supported by JSPS KAKENHI Grant Number  JP23H03438 and JP23K28128.

\bibliographystyle{IEEEtran} 
\bibliography{reference} 

\begin{thebibliography}{10}
\providecommand{\url}[1]{#1}
\csname url@samestyle\endcsname
\providecommand{\newblock}{\relax}
\providecommand{\bibinfo}[2]{#2}
\providecommand{\BIBentrySTDinterwordspacing}{\spaceskip=0pt\relax}
\providecommand{\BIBentryALTinterwordstretchfactor}{4}
\providecommand{\BIBentryALTinterwordspacing}{\spaceskip=\fontdimen2\font plus
\BIBentryALTinterwordstretchfactor\fontdimen3\font minus \fontdimen4\font\relax}
\providecommand{\BIBforeignlanguage}[2]{{%
\expandafter\ifx\csname l@#1\endcsname\relax
\typeout{** WARNING: IEEEtran.bst: No hyphenation pattern has been}%
\typeout{** loaded for the language `#1'. Using the pattern for}%
\typeout{** the default language instead.}%
\else
\language=\csname l@#1\endcsname
\fi
#2}}
\providecommand{\BIBdecl}{\relax}
\BIBdecl

\bibitem{Tsoli2011-jv}
A.~Tsoli and O.~C. Jenkins, ``Robot grasping for prosthetic applications,'' in \emph{Robotics Research}.\hskip 1em plus 0.5em minus 0.4em\relax Springer Berlin Heidelberg, 2011, pp. 1--12.

\bibitem{Rezazadeh2012-ol}
I.~M. Rezazadeh, M.~Firoozabadi, H.~Hu, and S.~M.~R. Golpayegani, ``Co-adaptive and affective human-machine interface for improving training performances of virtual myoelectric forearm prosthesis,'' \emph{IEEE Trans. Affect. Comput.}, vol.~3, no.~3, pp. 285--297, 2012.

\bibitem{Zhang2022-ld}
X.~Zhang \emph{et~al.}, ``\BIBforeignlanguage{en}{Domain adaptation with self-guided adaptive sampling strategy: Feature alignment for cross-user myoelectric pattern recognition},'' \emph{\BIBforeignlanguage{en}{IEEE Trans. Neural Syst. Rehabil. Eng.}}, vol.~30, pp. 1374--1383, May 2022.

\bibitem{Matsubara2011-pc}
T.~Matsubara, S.-H. Hyon, and J.~Morimoto, ``Learning and adaptation of a stylistic myoelectric interface: {EMG-based} robotic control with individual user differences,'' in \emph{Proc. IEEE Int. Conf. Robot. Biomim. (ROBIO)}, Dec. 2011, pp. 390--395.

\bibitem{Wu2023-bg}
D.~Wu, J.~Yang, and M.~Sawan, ``\BIBforeignlanguage{en}{Transfer learning on electromyography ({EMG}) tasks: Approaches and beyond},'' \emph{\BIBforeignlanguage{en}{IEEE Trans. Neural Syst. Rehabil. Eng.}}, vol.~31, pp. 3015--3034, Jul. 2023.

\bibitem{Zou2021-zy}
Y.~Zou and L.~Cheng, ``A transfer learning model for gesture recognition based on the deep features extracted by {CNN},'' \emph{IEEE Trans. Artif. Intell.}, vol.~2, no.~5, pp. 447--458, Oct. 2021.

\bibitem{Kim2020-ae}
K.-T. Kim, C.~Guan, and S.-W. Lee, ``\BIBforeignlanguage{en}{A subject-transfer framework based on single-trial {EMG} analysis using convolutional neural networks},'' \emph{\BIBforeignlanguage{en}{IEEE Trans. Neural Syst. Rehabil. Eng.}}, vol.~28, no.~1, pp. 94--103, Jan. 2020.

\bibitem{Bao2021-zq}
T.~Bao, S.~A.~R. Zaidi, S.~Xie, P.~Yang, and Z.-Q. Zhang, ``\BIBforeignlanguage{en}{Inter-subject domain adaptation for {CNN}-based wrist kinematics estimation using {sEMG}},'' \emph{\BIBforeignlanguage{en}{IEEE Trans. Neural Syst. Rehabil. Eng.}}, vol.~29, pp. 1068--1078, Jun. 2021.

\bibitem{Vidovic2016-bm}
M.~M.-C. Vidovic \emph{et~al.}, ``\BIBforeignlanguage{en}{Improving the robustness of myoelectric pattern recognition for upper limb prostheses by covariate shift adaptation},'' \emph{\BIBforeignlanguage{en}{IEEE Trans. Neural Syst. Rehabil. Eng.}}, vol.~24, no.~9, pp. 961--970, Sep. 2016.

\bibitem{Kanoga2021-ve}
S.~Kanoga, T.~Hoshino, and H.~Asoh, ``Semi-supervised style transfer mapping-based framework for {sEMG}-based pattern recognition with 1- or {2-DoF} forearm motions,'' \emph{Biomed. Signal Process. Control}, vol.~68, p. 102817, Jul. 2021.

\bibitem{Vidovic2014-jq}
M.~M.-C. Vidovic \emph{et~al.}, ``\BIBforeignlanguage{en}{Covariate shift adaptation in {EMG} pattern recognition for prosthetic device control},'' in \emph{\BIBforeignlanguage{en}{Proc. 36th Annu. Int. Conf. IEEE Eng. Med. Biol. Soc. (EMBC)}}, 2014, pp. 4370--4373.

\bibitem{Yoneda2023-lj}
S.~Yoneda and A.~Furui, ``Bayesian approach for adaptive {EMG} pattern classification via semi-supervised sequential learning,'' in \emph{Proc. IEEE Int. Conf. Syst. Man Cybern. (SMC)}, 2023, pp. 3310--3315.

\bibitem{Krasoulis2017-dr}
A.~Krasoulis, K.~Nazarpour, and S.~Vijayakumar, ``Use of regularized discriminant analysis improves myoelectric hand movement classification,'' in \emph{Proc. 8th Int. IEEE/EMBS Conf. Neural Eng. (NER)}, May 2017, pp. 395--398.

\end{thebibliography}

\end{document}